\documentclass[aps,pra,twocolumn]{revtex4}
\usepackage{graphicx}

\begin{document}

\title{Scalable Generation of Graph-State Entanglement through Realistic Linear
Optics}
\author{T. P. Bodiya, L.-M. Duan}
\address{FOCUS Center and MCTP, Department of Physics, University of Michigan, Ann Arbor, MI
48109}

\begin{abstract}
We propose a scheme for efficient construction of graph states using
realistic linear optics, imperfect photon source and single-photon
detectors. For any many-body entanglement represented by tree graph states,
we prove that the overall preparation and detection efficiency scales only
polynomially with the size of the graph, no matter how small the
efficiencies for the photon source and the detectors.
\end{abstract}

\maketitle

Linear optics, combined with practical photon source and single-photon
detectors, has provided a powerful tool to test a number of quantum
information protocols \cite{1,2,3,4,4'}. In the linear optics
implementations, the post-selection technique based on photon detections
typically plays a critical role. As one scales up the system, the
post-selection technique generally leads to very inefficient (exponential)
scaling of the overall efficiency, which limits the implementation to small
systems. In the past few years, two approaches have been proposed to
circumvent this obstacle for different kinds of quantum information
processing. In the first approach, a remarkable proposal, generally referred
to as linear optics quantum computation, was first put forward by Knill,
Laflamme, and Milburn (KLM) \cite{5} and then improved by a number of others
\cite{6,7,8,9,10}. In this approach, one overcomes the scaling problem for
linear optics computation through quantum error correction by requiring the
efficiencies for the photon source and the detectors to attain a high
threshold value \cite{5}. The threshold efficiency has been improved
considerably in the past few years, with the most recent estimate about $%
99.4\%-99.7\%$ \cite{10}. In the second approach proposed in Ref. \cite{11},
one uses linear optics for implementation of scalable quantum communication
through the quantum repeater protocol \cite{12}. An extension of that
protocol can also efficiently prepare GHZ\ type of entanglement \cite{13}.
This approach overcomes the inefficient scaling through the
divide-and-conquer method. In this protocol, one does not have a threshold
requirement on the source and the detector efficiencies \cite{note1}, which
allows its implementation with the state-of-the-art photon detectors \cite
{14}.

In this paper, we propose an efficient scheme for quantum state engineering
with linear optics. We show how to generate many-qubit entanglement
represented by graph states. Graph states have been generally identified as
a useful resource for many quantum information protocols, including quantum
computation, communication, and fundamental test of quantum mechanics \cite
{15,16}. The main result from this work is twofold: first, we analyze the
effect of a polarization beam splitter (PBS) in the Hilbert subspace
postselected by the photon detections, and show that a single PBS actually
represents a powerful gate for generating graph states of arbitrary shapes.
Other types of gates have been proposed before for preparation of graph
states \cite{6,7,8,9,10}, among them the most efficient one up to now seems
to be the fusion gate \cite{8}. Compared with that one, we show that a PBS
is a more efficient gate in the sense that it does not waste any photons for
each gate operation \cite{note2}. This improvement is desirable as in
current experiments the number of photons is a precious resource. Second, a
more important result here is a proposed method for scalable generation and
detection of many-qubit entanglement represented by tree graph states, with
the latter having a number of applications in recent quantum information
protocols \cite{16,9,17}. Here, by ``scalable'', we mean the overall
efficiency for preparation of a large-scale entanglement with a tree-graph
structure scales nearly polynomially with the number of qubits. We have this
efficient scaling no matter how small the efficiencies for the photon source
and the detectors. The scheme here thus well fits the status of the current
experimental technology.

We assume to have an imperfect source of entangled photon pairs, which
generates states of the following general form
\begin{equation}
\rho _{s}=\left( 1-\eta _{s}\right) \rho _{\text{vac}}+\eta _{s}\left| \Psi
\right\rangle _{12}\left\langle \Psi \right| ,
\end{equation}
where $\left| \Psi \right\rangle _{12}=\left( |{HH}\rangle _{12}+|{VV}%
\rangle _{12}\right) /\sqrt{2}$ represents a photon pair entangled in the
polarization states $|{H}\rangle $, $|{V}\rangle $; $\rho _{\text{vac}}$
stands for the vacuum component with no photon for the modes 1 and 2; and $%
\eta _{s}$ is the source efficiency for producing the entangled photon pair.
In experiments, the entangled photon source is typically provided through
the process of spontaneous parametric down conversion (SPDC), where the
source efficiency $\eta _{s}$ is a small number \cite{1,2,3,4}. The pair
state (1) can also be generated from other experimental setups, such as from
decay of a single dipole (which could be a single atom, ion, or a quantum
dot) in free space or in a cavity \cite{18,19}, or from decay of an
collective excitation in an atomic ensemble \cite{14}. In these cases, one
mode of the entangled pair is typically represented by a matter qubit, which
can be transferred later to a photon qubit after a controllable delay.

Now we show any graph state in the Hilbert subspace postselected
by the photon detection can be generated from the pair states (1)
through a series of PBS\ gates. An n-qubit graph state is defined
as the co-eigenstate of $n$ independent stabilizer operators
$S_{i}=X_{i}\prod_{j}Z_{j}$, where $i$ denotes qubit $i$ (each
qubit is associated with a vertex of the graph), $j$ runs over all
the neighbors of the qubit $i$, and $X_{i}$, $Z_{i}$ are simply
the Pauli operators $\sigma _{x}$ and $\sigma _{z}$ for qubit $i$
\cite{15,16}. In a graph, the qubits $i$ and $j$ are called
neighbors if they are connected with an edge. The graph state
reduces to a cluster state if the corresponding graph is a
periodic lattice \cite{15}.

To show construction of the graph states, first we need to analyze the
effect of a PBS in the subspace postselected by the photon detection. For
linear optics quantum information, all the photon modes will be eventually
measured in an appropriate polarization basis by a single-photon detector.
We are interested only in the measurement outcomes with one photon
registered from each mode (its polarization can be arbitrary). So, by this
final measurement, one postselects a Hilbert subspace, which  we denote as $S
$. We only need to find out the state evolution in this ``physical''
subspace $S$, as the state component outside $S$ has no influence on the
final measurement of the polarization qubits.

For a PBS, it transfers the photon if it in $H$ polarization and reflects it
if it is in $V$ polarization. So, after the PBS, the photons from the two
incoming modes go to different sides (modes) if and only if both photons
have the same polarization, either $HH$ or $VV$. Otherwise, they will go to
the same sides with the other mode in the vacuum state, which is outside of
the ``physical'' subspace $S$. So, within the subspace $S$, the effect of a
PBS is to perform a projection on the input state, described by the
projector
\begin{equation}
P=\left| HH\right\rangle _{12}\left\langle HH\right| +\left| VV\right\rangle
_{12}\left\langle VV\right| .
\end{equation}
This projection is equivalent to a measurement of the operator $Z_{1}Z_{2}$
on the two input qubits 1 and 2, with the final state kept only under the
measurement outcome ``$+1$'' ($\left| HH\right\rangle _{12}$ and $\left|
VV\right\rangle _{12}$ are eigenstates of $Z_{1}Z_{2}$ \ with an eigenvalue
``$+1$''). So in the physical subspace $S$, a single PBS performs an
effective $Z_{1}Z_{2}$\ measurement gate with a success probability of $1/2$
(the probability to stay in the ``physical'' space $S$ after the PBS).

We start with two entangled pairs 1,2 and 3,4, each pair described by the
state (1). In the subspace $S$, the effective state is then given by $\left|
\Psi \right\rangle _{12}$, which can be transferred to a two-bit graph state
with a straightforward Hadamard gate on one of the qubits. So, for the pairs
1,2 and 3,4, we can assume them to have the stabilizer operators $X_{1}Z_{2},
$ $X_{2}Z_{1}$ and $X_{3}Z_{4},$ $X_{4}Z_{3}$, respectively. If the qubits 2
and 3 pass through a PBS, the effective output state in the subspace $S$ is
then stabilized by the operators $Z_{2}Z_{3},$ $X_{2}X_{3}Z_{1}Z_{4},$ $%
X_{2}Z_{1},$ $X_{4}Z_{3}$ ($X_{2}X_{3}Z_{1}Z_{4}$ is a multiplication of the
previous stabilizers $X_{2}Z_{1}$ and $X_{3}Z_{4}$, and it remains unchanged
after the PBS as it commutes with the effective measurement gate $Z_{2}Z_{3}$%
). With a straightforward Hadamard gate $X_{3}\leftrightarrow Z_{3}$
(implemented with a half-wave plate), the above four stabilizers transform
to the standard stabilizers for the 4-bit star-shape graph state as shown in
Fig. 1.

For convenience, we call a combination of a PBS and a single-bit Hadamard
operation as the PBS gate (see Fig. 1). An extension of the above
construction yields the following important result: the PBS gate always
joins two pieces of graphs, independent of the shapes of the initial pieces.
This result can be generally proven as follows: if we start with two pieces
of graph states $G_{1}$ and $G_{2}$, with $n$ and $m$ qubits, respectively.
The stabilizers associated with the qubits $i_{1}$ and $i_{2}$ are given by $%
S_{i_{l}}=X_{i_{l}}\prod_{j_{l}\in N\left( i_{l}\right) }Z_{j_{l}},$ $\left(
l=1,2\right) $, where $i_{l}$ is an arbitrary vertex of the graph $G_{l}$
and $N\left( i_{l}\right) $ denotes all the neighbors of the qubit $i_{l}$
in the graph $G_{l}$. After a PBS gate on the qubits $i_{1}$ and $i_{2}$,
the stabilizers $S_{i_{2}}$ and $S_{i_{1}}$ are replaced by $%
S_{i_{2}}^{\prime }=X_{i_{2}}Z_{i_{1}}$ and $S_{i_{1}}^{\prime
}=X_{i_{1}}Z_{i_{2}}\prod_{j_{1}\in N\left( i_{1}\right)
}Z_{j_{1}}\prod_{j_{2}\in N\left( i_{2}\right) }Z_{j_{2}}$. All the other
stabilizers of the initial graphs $G_{1}$ and $G_{2}$ remain unchanged after
the gate. One can immediately see that the effective output state of the PBS
gate is still a graph state which combines the two initial graphs $G_{1}$
and $G_{2}$, with $i_{2}$ attached to $i_{1},$ and $i_{1}$ attached to $i_{2}
$ and all their initial neighbors in the graphs $G_{1}$ and $G_{2}$ (see
Fig. 1[c]).

With the above result, it becomes possible to construct any shapes
of the graph states with a series of PBS gates. In Figures 1 and
2, we illustrate the method by constructing the states represented
respectively by a tree graph and by a two-dimensional graph with
many loops. This construction method is efficient in the sense
that no photon is wasted during the state preparation. Starting
with $n$ entangled pairs, we get graph states of $2n$ qubits with
various shapes.
\begin{figure}[tbp]
\includegraphics[height=4cm,width=8cm]{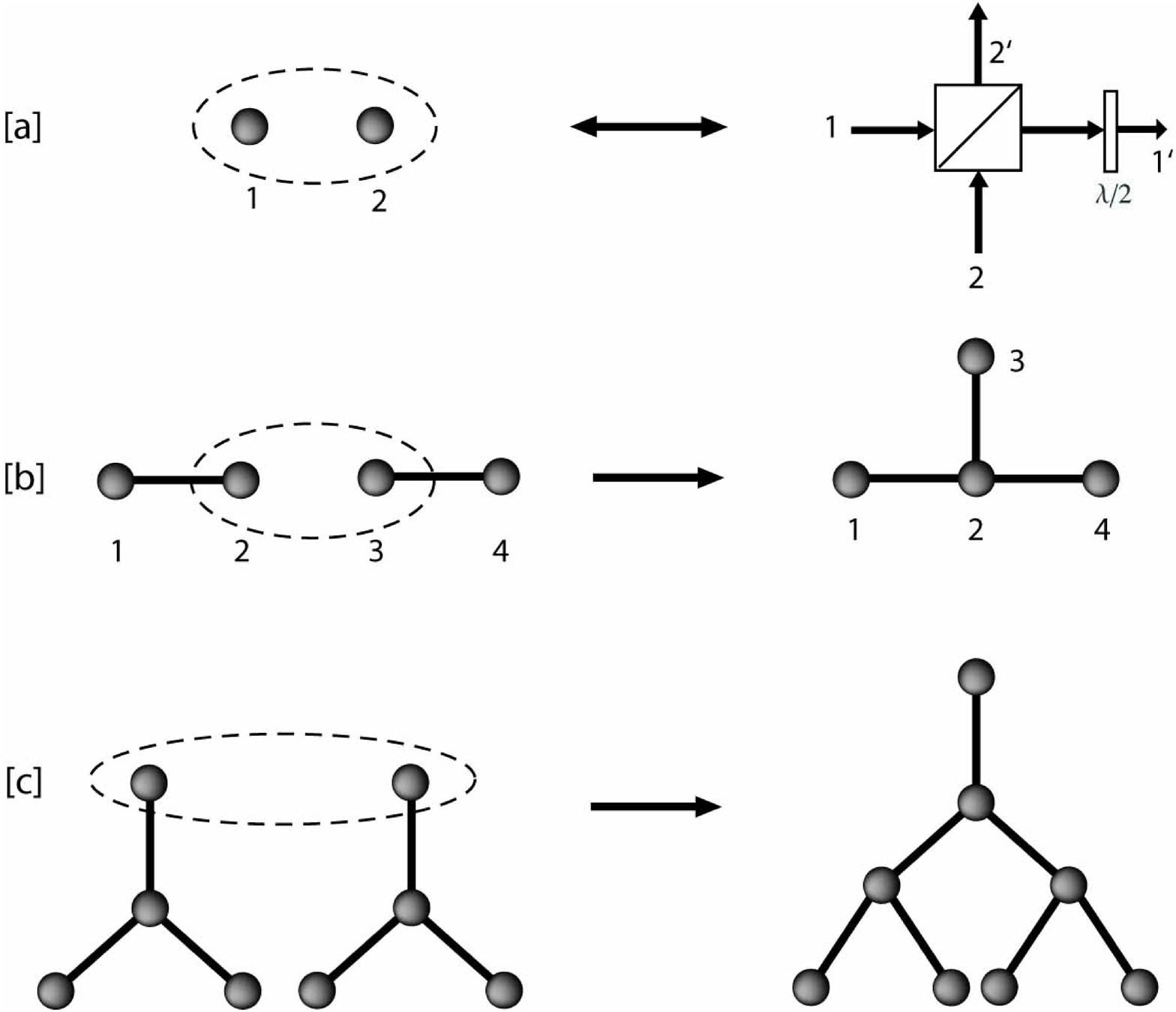}
\caption{[a]: The representation of the PBS gate, which consists
of a polarization beam splitter and a half-wave plate (for a
Hardamard operation on one mode). [b] and [c]: Illustration of
using the PBS gates to generater tree graph states. It is obvious
that tree graphs of any shapes can be generated with this method.
} \label{fig:Figure1V2}
\end{figure}
\begin{figure}[tbp]
\includegraphics[height=4cm,width=8cm]{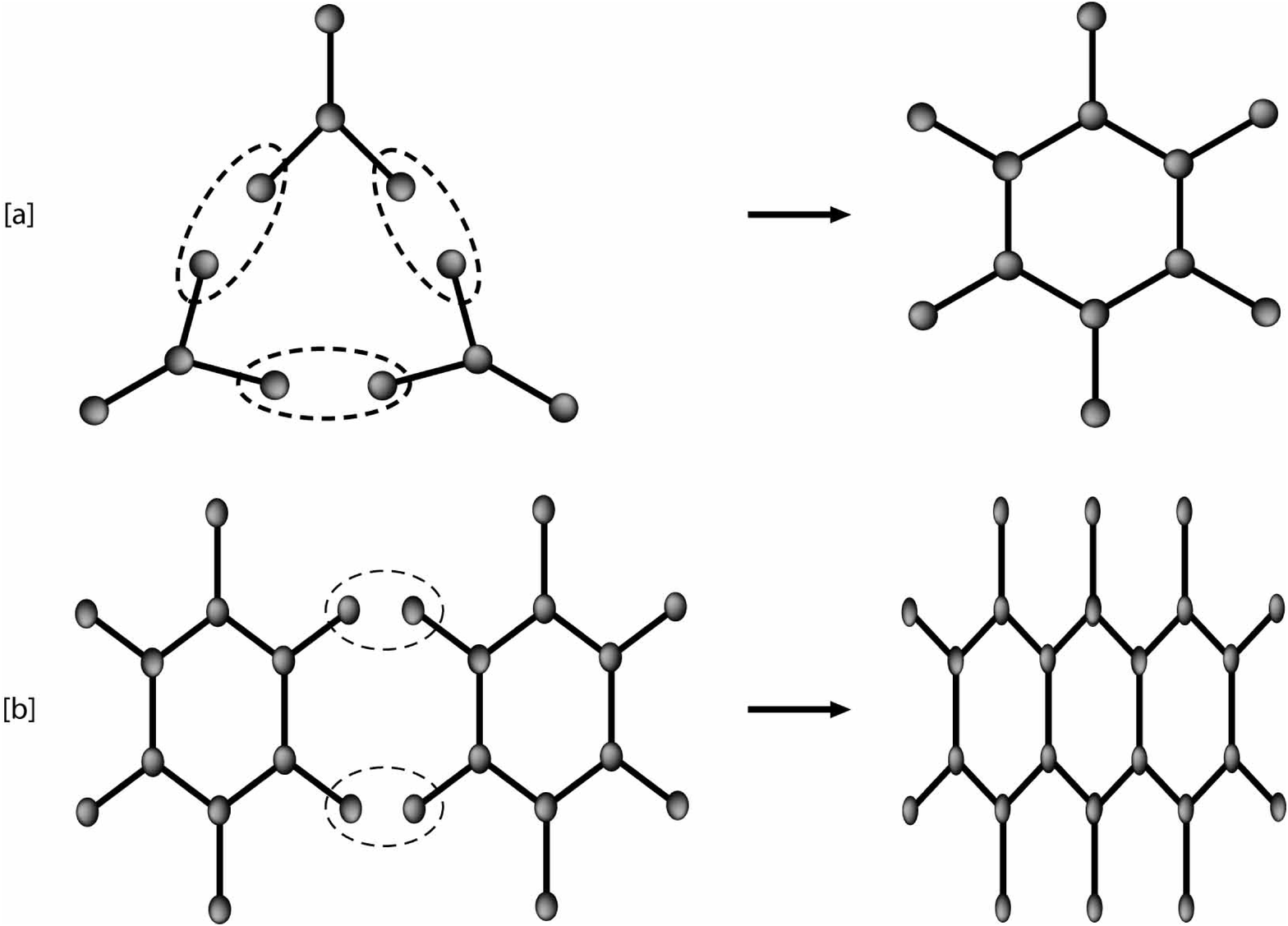}
\caption{[a] and [b]: Illustration of using the PBS gates to
generate 2-dimensional graphs states.  } \label{fig:Figure2V2}
\end{figure}

If we assume both the photon source and the detectors have large
inefficiencies, in general we still have inefficient (exponential) scaling
for construction of large-scale graph states, even after the above
improvement \cite{note3}. If one wants to generate an $n$-qubit graph-state
entanglement, one needs to consume $n/2$ imperfect entangled pairs
represented by the state (1) and detect $n$ photon modes at the end. So,
there is a factor of $\eta _{d}^{n}\eta _{s}^{n/2}$ in the preparation
efficiency, where $\eta _{d}$ is the efficiency for each individual
detector. If we need $m$ $\left( m\leq n/2\right) $\ PBS\ gates to arrive at
such a graph state, there is also additional factor of $\left( 1/2\right)
^{m}$ in the preparation efficiency associated with the intrinsic gate
success probability (to stay in the subspace $S$). In the case of a small
source efficiency $\eta _{s}$ (such as for the SPDC experiments), the
preparation efficiency goes down pretty quickly with the size of the state,
which limits the current implementation to a small number of qubits \cite
{1,2,3,4,4'}. In the following, we will show that for a subclass of
graph-state entanglement, that is, for any $n$-qubit entanglements
represented by tree graph states, we can prepare and detect them very
efficiently with a polynomial scaling over the state size.

The achievement of this efficient scaling is based on a combination of the
ideas of the divide-and-conquer (quantum repeater) protocol and the
postselection measurements. We note that for applications of graph states in
linear optics quantum information, each photon mode needs to be eventually
measured in some polarization basis. This suggests that the whole protocol
can be divided into two logical steps: the graph state preparation and the
application measurements. For the second step, measurement of each photon
mode has a finite failure probability, where instead of getting the photon's
polarization, one does not register any photon. To boost the efficiency of
the whole protocol, it is better to sort out and discard these failure
events as soon as possible. In this spirit, we can try to apply the
applications measurements on some individual qubits before we finish the
first logic step of the graph-state preparation. We measure the qubits as
soon as we do not need to apply the PBS\ gates on those qubits any more.
When we register a failure event, we immediately discard the qubits that are
influenced by the failure event, and restart the state preparation for that
segment.

Figure 3 illustrates how such an idea works for preparation and detection of
a tree-graph state. We start with two pairs (1,2) and (3,4), with the pair
state described by Eq. (1). As we do not need to apply the PBS gates on the
qubits 1 and 4 in the following steps, we immediately measure them in the
polarization basis chosen according to the targeted application protocol.
The measurement on the qubit 1 (or 4) succeeds with a probability $%
p_{0}=\eta _{s}\eta _{d}$, and upon a success, the vacuum component in the
imperfect state (1) is eliminated. If we fail for the measurement on the
qubit 1, we only need to re-prepare the state for the pair (1,2), with the
pair (3,4) intact \cite{note4}. With in average $1/p_{0}$ trials, we succeed
to get a good effective state (without the vacuum component) for the pair
(1,2). Parallel to this effort, we also get a good state for the pair (3,4),
similarly with $1/p_{0}$ trials. Then, we continue with the connection of
the qubits 2 and 3 through a PBS gate, and after the connection, we
immediately measure the qubit 2 as we only need keep the qubit 3 for the
next step of connection. This process is continued until we get an effective
tree-graph state with a desired number of qubits.
\begin{figure}[tbp]
\includegraphics[height=4cm,width=8cm]{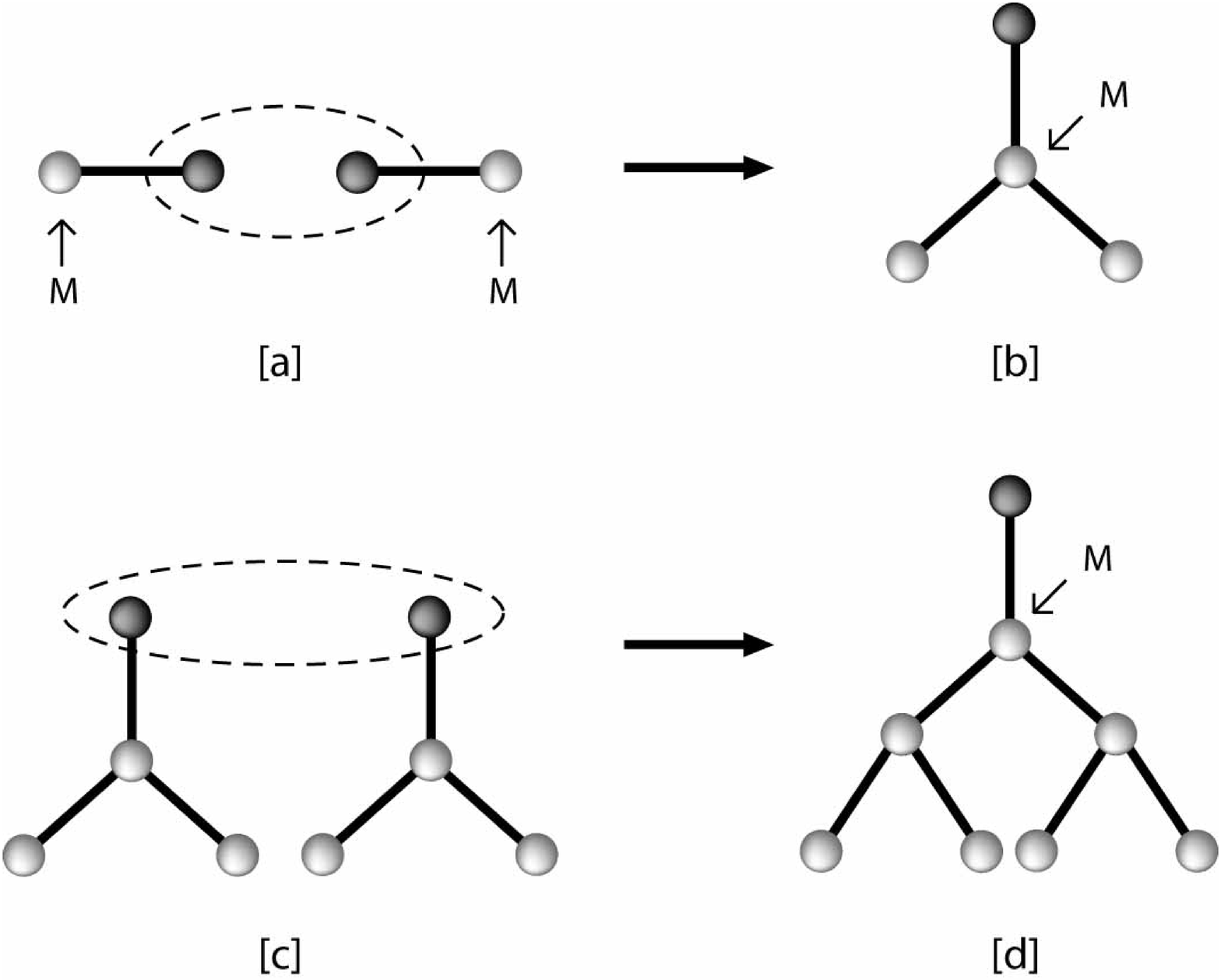}
\caption{Efficient construction of tree-graph states. White
circles represent qubits that have have been measured in
appropriate polarization bases, and black circles represent the
connection qubits (unmeasured) which enable the next-step
connection. [a] Before connection of the two center qubits, the
two edge qubits have been measured. [b] After connection, we
immediately measure one of the connection qubits, and leave the
other one for the next step connection as shown in [c]. [c-d]
Repeat the process of connection-and-measurement for construction
of larger graphs. } \label{fig:Figure3V2}
\end{figure}

To figure out the overall efficiency for generation of this graph-state
entanglement, we need to specify the recursion relations for each step of
connection. For each connection, the number of qubits is doubled. For the $m$%
th connection, the effective state before connection can be written as $\rho
_{2n}^{a}=\rho _{n}\otimes \rho _{n}$, where $\rho _{n}$ is the state of a
segment which has $n=2^{m}$ qubits. The segment state can be expressed as $%
\rho _{n}=a_{m-1}\rho _{g}+\left( 1-a_{m-1}\right) \rho _{\text{vac}}$,
where $\rho _{g}$ denotes the effective $n$-qubit tree graph state and $\rho
_{\text{vac}}$ represents the vacuum component where the connection qubit of
the graph is in the vacuum state. For the 1st connection (the connection of
the pair (1,2) and (3,4)), as the vacuum component has been eliminated by
the measurement on the qubits 1 and 4, we have $a_{0}=1$. After the $m$th
connection, we immediately measure one of the two connection qubits (the
other one is kept as the connection qubit for the next step). The success
probability for this measurement is given by
\begin{equation}
p_{m}=\eta _{d}\left[ a_{m-1}^{2}/2+a_{m-1}^{2}\left( 2-\eta _{d}\right)
/4+a_{m-1}\left( 1-a_{m-1}\right) \right] ,
\end{equation}
where we have assumed the detector cannot distinguish the single-photon and
two-photon counts, as it is the case in practice. Upon a success of this
measurement, the effective state for the $2n$ qubits becomes $\rho
_{2n}=a_{m}\rho _{g}+\left( 1-a_{m}\right) \rho _{\text{vac}}$, where $\rho
_{g}$ and $\rho _{\text{vac}}$ have the same meaning as before except that
they are for $2n$ qubits now, and the coefficient $a_{m}$ is given by the
recursion relation $a_{m}=2a_{m-1}/\left( 4-\eta _{d}a_{m-1}\right) $.
Together with $a_{0}=1$, this recursion relation yields
\begin{equation}
a_{m}=\left[ 2^{m}\left( 1-\eta _{d}/2\right) +\eta _{d}/2\right] ^{-1}.
\end{equation}

To prepare and confirm an $n=2^{m}$ qubit entanglement represented by the
tree graph state, the overall efficiency of the scheme can be characterized
by the total preparation time $T$. From the above recursion relations, one
can find that
\begin{eqnarray}
T &=&t_{0}\left( \eta _{d}a_{m-1}\right) ^{-1}\prod_{i=0}^{m-1}\left(
1/p_{i}\right)   \nonumber \\
&\approx &t_{0}\left( \eta _{s}\eta _{d}\right) ^{-1}n^{\left[ \left( \log
_{2}n-1\right) /2+\log _{2}\left( 1/\eta _{d}-1/2\right) \right] },
\end{eqnarray}
where the approximation is valid when $\eta _{d}/2\ll n$, and we have
assumed that the two segments of graphs states before each connection can be
prepared in parallel. The $t_{0}$\ in $T$ denotes the time to generate the
imperfect pair (1), which is basically the inverse of the pulse repetition
rate in the SPDC experiment \cite{1,2,3,4}. On can see that $T$ scales
nearly polynomially with the size $n$ of the final graph state, and such a
scaling holds for any positive source efficiency $\eta _{s}$ and detector
efficiency $\eta _{d}$.

Before ending the manuscript, let us briefly mention some
practical implications of this method. If we take the source
efficiency $\eta _{s}\sim 1\%$ and the detector efficiency $\eta
_{d}\sim 70\%$, as it is typical for current experiments
\cite{1,2,3,4}, we get $T/t_{0}\sim 1.8\times 10^{8}$ for
preparation of a graph state of $128$ qubits. If the pulse
repetition rate is $80$ MHz (the value from the SPDC experiment
\cite{1,2,3,4}), the total preparation time $T$ will be about $2$
seconds, which is still pretty reasonable. If we do not use this
divide-and conquer technique, the total time will be given by
$T/t_{0}=\eta _{s}^{-64}\eta _{d}^{-128}2^{128/2-1}\sim 10^{167}$
for entanglement of $128$ qubits, which is terribly long. We
should also mention that we neglect in this paper other sources of
noise except the photon source and detector inefficiencies. This
is well justified as other noise is typically much smaller in
magnitudes. For instance, the dark count probability within the
pulse interval ($\sim 1/80$ $\mu s$) is below $10^{-4}$, which
indicates such a noise is negligible for entanglement of hundreds
of qubits.

In summary, we have shown that a single polarization beam splitter acts as a
powerful gate in the physical subspace postselected by the photon detection.
Such a gate is very efficient for generating graph state of any shapes. In
particular, we have shown for preparation and detection of tree graph
states, the required resources scale only polynomially with size of the
graph. This result opens up a prospect to generate large scale entanglement
with the state-of the-art technology, while such entanglement is useful for
implementation of various quantum information protocols.

This work was supported by the NSF awards (0431476), the ARDA under ARO
contracts, and the A. P. Sloan Fellowship.


\begin{thebibliography}{99}
\bibitem{note1}  To achieve efficient scaling for quantum communication
through the divide-and-conquer method, the photon source and detector
efficiencies can be any positive number, which is equivalent to say that the
threshold efficiency is $0\%$ for this particular noise model.
\bibitem{note2}  Note that in the type-I or type-II fusion gate \cite{8},
one detects one or two output modes, which means that one has to discard one
or more photons for each gate operation.
\bibitem{note3}  If we assume perfect detectors and photon source, we can
construct any graph state with an efficient scaling through the PBS gate
with the method of Refs. \cite{7,8}. In this work, however, we would like to
insist on a more realistic assumption with large inefficiencies for both of
the photon source and the detectors.
\bibitem{note4}  We assumed here that we can keep the pair (3,4) for a
little while to wait for the success of the measurement on the qubit 1. Such
a memory (a controllable delay) is an essential requirement for scalability
of all linear optics quantum information protocols, see \cite{4',14,19} for
some experiments in this direction.
\end{thebibliography}
\end{document}